\documentclass[prl,twocolumn,lengthcheck]{revtex4b5}

\usepackage{graphicx}

\begin{document}

\title{An adaptive inelastic magnetic mirror for Bose-Einstein condensates}

\author{A. S. Arnold}
\email{a.arnold@phys.strath.ac.uk}
\author{C. MacCormick}
\author{M. G. Boshier}
\email{m.g.boshier@sussex.ac.uk}
\affiliation{Sussex Centre for
Optical and Atomic Physics, University of Sussex, Brighton BN1
9QH, UK}

\begin{abstract}
We report the reflection and focussing of a Bose-Einstein
condensate by a new pulsed magnetic mirror.  The mirror is
adaptive, inelastic, and of extremely high optical quality.
The deviations from specularity are less than
$0.5\,\mbox{mrad}$ rms, making this the best atomic mirror
demonstrated to date.  We have also used the mirror to
realize the analog of a beam-expander, producing an
ultra-cold collimated fountain of matter waves.
\end{abstract}


\maketitle

A major aim of the field of atom optics is to build analogs of
mirrors, lenses, and waveguides for manipulating ultra-cold atoms
in applications ranging from high resolution lithography to
ultra-sensitive atom interferometry. To this end, static magnetic
fields and the optical dipole potential have already been used to
both reflect and focus
\cite{Roach,Saba,Drndic,Lau1,Kasev1,Amin,Hinds} conventional cold
atomic clouds, and laser-cooled neutral atoms have also been
focused in both 1D \cite{Marec} and 3D \cite{mon1} using pulsed
magnetic fields.  Since a gaseous Bose-Einstein condensate (BEC)
\cite{And1} is the ultimate source of coherent atoms, it is
essential to develop high-quality atom-optical elements which can
be applied to BECs.  A first step in this direction was the recent
demonstration of a flat mirror based on the optical dipole force
\cite{Bongs}.

In this Letter we show that a different kind of atomic mirror
\cite{ArnProc}, based on pulsed magnetic fields \cite{ArnProc,
Bloch2}, can be used to manipulate Bose condensates. This new
mirror has several attractive features: it can be inelastic, it
permits adjustable three-dimensional focusing, and it has
extremely high optical quality. In addition to demonstrations of
simple reflection and focussing, these properties also allow us to
realize a BEC ``beam expander'' which produces an extremely cold,
collimated beam of matter waves. Finally, we show that the
evolution of the condensate in these experiments is in good
agreement with theory.

All magnetic atom-optical elements make use of the Stern-Gerlach
potential $U=-\mu B$ experienced by an atom moving adiabatically
in a magnetic field of magnitude $B$ (with $\mu$ being the
component of the atomic magnetic moment parallel to the field).
For weak-field seeking atoms, any field for which $B$ increases in
the direction of the initial atomic velocity will act as a simple
mirror. If the atoms are to be focussed as well as reflected, then
$B$ must in addition have positive curvature in the appropriate
directions. Other desirable properties of a condensate mirror are
equipotential surfaces with sufficient smoothness to preserve
coherence, and compatibility with the ultra-high vacuum
environment of a typical BEC apparatus. Since these conditions
must all be satisfied by the magnetic trap in which the condensate
is formed, the trap field itself provides an ideal starting point
for the construction of a mirror. Particularly important here is
the fact that all magnetic elements in the trap are a relatively
large distance $d\gtrsim20\,\mbox{mm}$ from the atoms. Thus,
because curvature in the magnetic field scales as $d^{-3},$
microscopic corrugations are drastically reduced compared to
mirrors in which atoms make a close $(d<100\,\mu\mbox{m})$
approach to the surface. Such corrugations, and their optical
analogs, presently limit the quality of existing atomic mirrors
\cite{Hinds, Asp} and so large-scale mirrors might provide an
easier route to fully coherent atomic manipulation. Also, with
this mirror the condensate is focused in three dimensions with a
single magnetic pulse, in contrast to previous work with cold
atoms which has either been restricted to one dimension
\cite{Marec} or required the judicious application of two magnetic
pulses \cite{mon1}.

\begin{figure}
\includegraphics{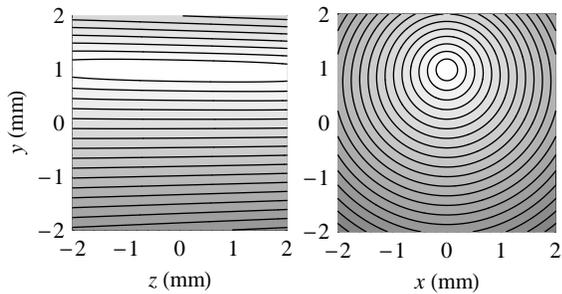}
\caption{Contours of the magnitude of the magnetic mirror field
for the case $y_{c}=1\,\mbox{mm}$. Gravity acts downwards.}
\label{conts}
\end{figure}

In our experiment the magnetic mirror is formed by combining a
horizontal Ioffe-Pritchard \cite{Pri} magnetic trap with a uniform
vertical magnetic field $B_{c}$.   We use a co-ordinate system
where the symmetry axis of the trap is in the $z$ direction,
gravity acts in the $-y$ direction and the origin is at the center
of the magnetic trap.  The total magnetic mirror field to
second-order is then
\begin{eqnarray}
 \textbf{B}(\textbf{r})&=&\{0,B_{c},B_{0}\} + B' \{x,-y,0\} \nonumber \\
           &&+B''/2\{-x z,-y z,z^2-(x^2+y^2)/2\}.
 \label{IPfield}
\end{eqnarray}
The parameters $B_{0},$ $B',$ and $B''$ are respectively the bias
field, gradient, and curvature of the trap.  One can show that the
magnitude of this field (Fig.~\ref{conts}) is essentially
parabolic in the axial $(z)$ direction, but hyperbolic in the
offset radial co-ordinate $r=\sqrt{x^2+(y-y_{c})^2}$ and thus
effectively linear for radii $r>\sqrt{2}B_0/B'.$  The potential
therefore closely approximates that of a horizontal cylindrical
mirror.  The control field $B_{c}$ allows us to shift the minimum
of the potential vertically from $y = 0$ to $y_{c}=B_{c}/B'$,
moving the center of curvature of the mirror to $y_{c}$ and making
the radius of curvature at a particular height $-y$ below the trap
center equal to $y + y_{c}$.  A condensate falling under gravity
can be reflected by a brief pulse of the magnetic mirror field.
During the pulse the atoms experience a potential that is weakly
confining in the $z$ direction, strongly confining (with strength
dependent on $B_c$) in the $x$ direction and almost linear in the
$y$ coordinate.  The linear variation of the potential with height
results simply in a change in the center-of-mass motion (i.e.
bouncing), whereas the parabolic dependence in the two horizontal
directions focuses the condensate. The field pulses can be
repeated at appropriate intervals to produce multiple reflections.

The BECs used in our experiments are produced as follows. About
$10^{9}$ $^{87}$Rb atoms are collected at the low-pressure end of
a double magneto-optical trap (MOT) \cite{gib2} system. Three
simple extended-cavity diode lasers \cite{arn} provide the
necessary $780\,\mbox{nm}$ $(5s\,^{2}S_{1/2}\rightarrow
5p\,^{2}P_{3/2})$ MOT trapping and repumping light. For both MOTs
the trap laser light is detuned $\Delta=-13\,\mbox{MHz}$ (i.e.
red) of the $F=2\rightarrow F'=3$ transition, whilst the repumping
light is resonant with the $F=1\rightarrow F'=2$ line. The atoms
are cooled to $40\,\mu\mbox{K}$ in optical molasses
$(\Delta=-35\,\mbox{MHz}),$ optically pumped on the
$F=2\rightarrow F'=2$ transition into the
$|F,m_{F}\rangle=|2,2\rangle$ state, and then transferred into the
Ioffe-Pritchard magnetic trap. The trap coils are formed from
water-cooled $3\,\mbox{mm}$ o.d. copper tubing: a 9-turn cuboidal
baseball coil with average side dimensions
$45\times45\times55\,\mbox{mm}^{3},$ and two three-turn
$64\times101\,\mbox{mm}^{2}$ rectangular bias coils.  With a
typical bias field of $1\,\mbox{G},$ the axial and radial magnetic
trapping frequencies were measured to be $\nu_{z}=10\,\mbox{Hz}$
and $\nu_{r}=223\,\mbox{Hz}$ respectively at a current of
$220\,\mbox{A}.$ The trap lifetime was typically $70\,\mbox{s}.$
After a $32\,\mbox{s}$ RF evaporative cooling ramp pure
condensates are reproducibly obtained with more than $10^{5}$
atoms. The atoms are imaged in absorption using an
$8\,\mu\mbox{s}$ pulse of near-resonant light from a beam which
propagates horizontally at an angle of $60^{\circ}$ to the $z$
axis of the magnetic trap.  We refer to the resulting horizontal
axis in our images as $z'$. The imaging system has a magnification
of $0.80.$ Comparison of the absorption images with a Thomas-Fermi
model \cite{Cast1} shows that distortion of the probe beam by the
initially optically dense BEC \cite{Andr1} becomes negligible
after $10\,\mbox{ms}$ of expansion.

\begin{figure*}
\includegraphics{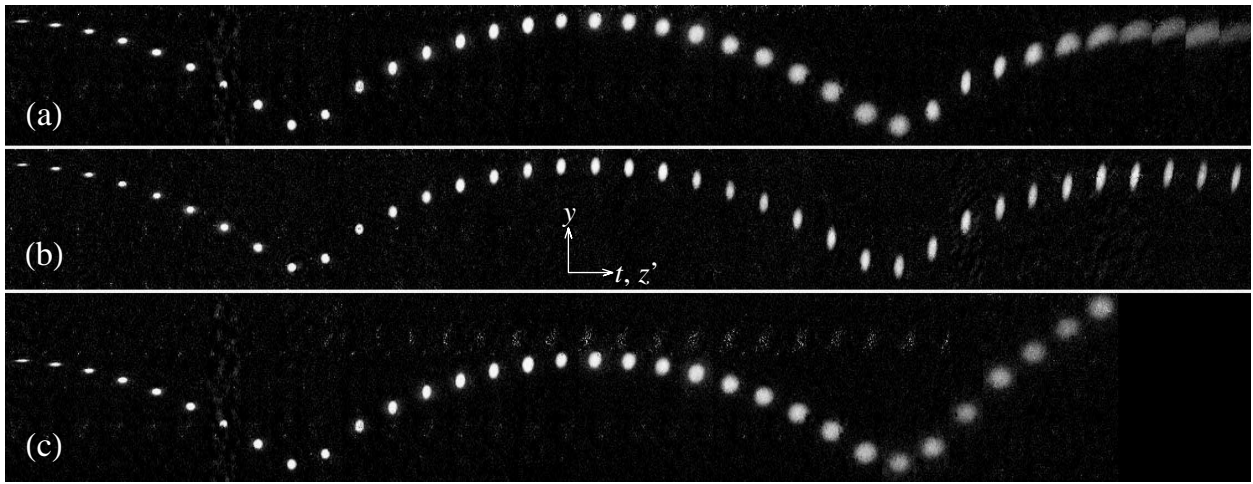}
\caption{Sequences of experimental absorption images taken at
times  $t=2,4,6,...,74\,\mbox{ms}$ for control fields (a) $B_c=0$
and, (b) $B_c=70\,\mbox{G}$. Each individual image in a sequence
is $0.5\,\mbox{mm}$ wide (the $z'$ direction) and $2.0\,\mbox{mm}$
high (the $y$ direction). In the image sequence (c) the control
field is $B_c=0$ during the first bounce, and the control field
and pulse duration during the second bounce are chosen to
collimate and launch the BEC. In this sequence the individual
images have dimensions $0.5\times2.7\,\mbox{mm}^{2}$.}
 \label{BounceSeq}
\end{figure*}

Fig.~\ref{BounceSeq} (a) and (b) are typical sequences of
absorption images showing reflection and focussing of the
condensate by mirrors of small and large radii of curvature
respectively. The magnetic field, with a current of
$155\,\mbox{A}$ in the magnetic trap coils, was pulsed on for
$5\,\mbox{ms}$ every $35\,\mbox{ms}$. This pulse duration was
chosen to give an elastic bounce for our magnetic acceleration $
\mu B' /m \approx 6g.$ In these sequences we have suppressed a
horizontal motion of the condensate, discussed below, to
illustrate the bouncing more clearly and focus attention on the
evolution of the condensate shape.

The rapid growth in the condensate width seen in
Fig.~\ref{BounceSeq} (a) is a consequence of the instability of
the cavity formed by the magnetic mirror and gravity. A
gravity-cavity is unstable, meaning that the atomic trajectories
walk out of the cavity, when the the radius of curvature $R$ of
the mirror is less than twice the release height $h$.
\cite{wal,Hinds,HughesNew}. This is the case for
Fig.~\ref{BounceSeq} (a), where the radius of curvature in the
$xy$ plane $R_{xy} = h \approx 1.4\,\mbox{mm}$.  The corresponding
classical motion is easily visualized in this case.  Since the
gravitational potential energy is much larger than the initial
kinetic energy of the condensate, all atomic trajectories strike
the mirror at an angle very close to vertical. The reflected
trajectories then all cross the mirror axis at a focal point
$\approx0.5 R_{xy}$ above the mirror surface on their way to
turning points at height $h$. The condensate will therefore be
strongly focused through this focal point after each bounce, as
can be can be clearly seen after the second bounce in
Fig.~\ref{BounceSeq} (a), even though our viewpoint is at an angle
of $60^{\circ}$ to the $z$ axis.  In contrast Fig.~\ref{BounceSeq}
(b) corresponds to the stable case $R_{xy} = 5h$.  The evolution
here is similar to that expected for bouncing on a flat mirror.

A complete theoretical analysis of our experiment would require a
numerical integration of the Gross-Pitaevskii equation on a large
mesh in three dimensions, a rather formidable task.  We have
therefore developed two simpler theoretical models of the bounce
dynamics. The first extends the work of Castin and Dum
\cite{Cast1} who showed that a BEC, in the Thomas-Fermi regime and
confined in a time-dependent parabolic potential
 $U(\textbf{r},t)=\frac{m}{2}\sum_{j=1}^{3}{\omega_{j}(t)^2{r_{j}}^2},$
has an atomic spatial distribution which obeys the simple scaling
law:
 \begin{equation}
 n(\textbf{r},t)\!=\!\max\!\left\{\frac{n_0}{\lambda_{1}\lambda_{2}\lambda_{3}}
 \!\!\left(1-\sum_{j=1}^{3}{\frac{{r_{j}}^{2}}{{A_{j}}^2 {\lambda_{j}}^2}}
 \right)\!\!,0\right\}
 \label{TF1}
 \end{equation}
where $\lambda_{j}(t)$ is the solution of
 \begin{equation}
 \frac{d^2\lambda_{j}(t)}{dt^2}+\omega_{j}(t)^2\lambda_{j}(t)-
     \frac{\omega_{j}(0)^2}{\prod_{i=1}^{3}{\lambda_{i}(t)^{1+\delta_{ij}}}}=0,
 \label{TF2}
 \end{equation}
 $\lambda_{j}(0)=1$ and the initial Thomas-Fermi radii are $A_{j}.$
It can be shown \cite{ArnNext} that if the potential is parabolic
about the time-varying center
 $\textbf{r}_c(t)$ of the BEC, i.e. if
 $U(\textbf{r},t)=\frac{m}{2}\sum_{j=1}^{3}{\omega_{j}(t)^2
 (r_{j}-{r_{c}}_j(t))^2},$ then the \textit{same}
scaling law applies for the BEC if one subtracts out the overall
center-of-mass motion and uses the magnetic curvatures acting at
the condensate center.  This model is analytic, and therefore fast
to compute, and it includes the effects of mean-field repulsion.
In order to verify that this locally harmonic potential
approximation is valid for our mirror, we developed a second
model, based on a Monte Carlo simulation of the classical
dynamics.  The simulation computes the classical trajectories of
$10^5$ atoms with initial positions and velocities chosen at
random so as to give the Thomas-Fermi result for the free
expansion (Eq.~\ref{TF1}). In this model the mirror potential can
be treated exactly, but atomic interactions after the initial
mean-field energy driven expansion are neglected. Both theoretical
models make the reasonable approximation that diffraction is
negligible, and both use values of the parameters characterizing
our mirror ($B_0,B',B'',B_c$) measured in other experiments.

\begin{figure}
\includegraphics{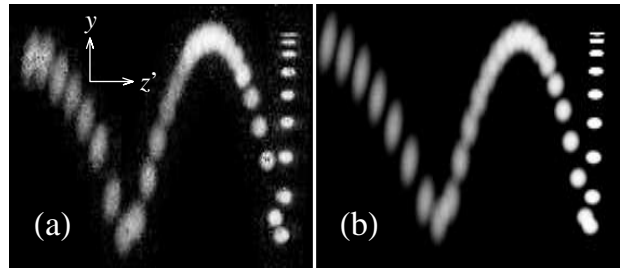}
\caption{(a) Superimposed experimental images for times
$t=2,4,6,...,68\,\mbox{ms}$ showing the full center-of-mass motion
for a mirror with $B_c=70\,\mbox{G}$ (i.e.
Fig.~\ref{BounceSeq}(b)). The image is in the $z'y$ plane, with
size $2.2\times1.9\,\mbox{mm}^{2}.$ (b) Monte Carlo simulation of
the condensate evolution for the
same conditions.}%
\label{ExptMC}
\end{figure}

Fig.~\ref{ExptMC} shows that the Monte Carlo simulation is in
excellent agreement with the experimental data of
Fig.~\ref{BounceSeq} (b). The horizontal motion seen here (and
suppressed in Fig.~\ref{BounceSeq}) is partly due to a tilt in the
magnetic trap with respect to gravity, and partly due to the fact
that for our coil orientation the magnetic potentials have an
increasingly large slope in the $z$ direction as $y$ decreases
(see Fig.~\ref{conts}). This slope also causes the condensate to
rotate in the $zy$ plane during a bounce. The analytic
Thomas-Fermi calculation yields very similar results, except that
the rotation of the condensate is not reproduced because this
model assumes a perfectly harmonic trap.

\begin{figure}
\includegraphics{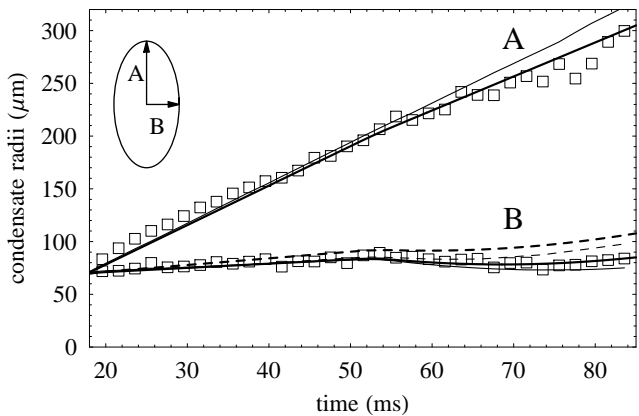}
\caption{Squares show the experimentally-determined BEC radii
(measured as shown in the inset) as a function of time for a
condensate bouncing on a mirror with $R_{xy}=5h$
(Fig.~\ref{BounceSeq} (b)). The bold and light curves show the
predictions of the Thomas-Fermi and Monte Carlo models
respectively.  The two dashed curves show the corresponding
predictions for the case of a mirror with rms roughness of
$0.5\,\mbox{mrad}$.}%
\label{BecRadii}
\end{figure}

To make a more quantitative comparison of the experimental data
with these models, we fitted both the Monte Carlo simulation and
the experimental absorption images to a Thomas-Fermi column
density to obtain the condensate radii (see inset in
Fig.~\ref{BecRadii}).  These two quantities can also be obtained
directly from our analytic Thomas-Fermi model.  The comparison
between experiment and the two models is shown in
Fig.~\ref{BecRadii}.  The two models are in good agreement with
each other, confirming the validity of their respective
approximations, and both are in good agreement with the measured
condensate radii.

The horizontal size of the bouncing condensate is a sensitive
probe of the optical quality of the mirror because any
corrugations or other aberration will increase the size of the
condensate on each bounce.  We therefore repeated the two
simulations, adding the effects of an rms mirror roughness of
$0.5\,\mbox{mrad}$ to the condensate on each bounce. The results
are shown as the dashed curves in Fig.~\ref{BecRadii}. The
discrepancy between these curves and the experimental data implies
that any deviations from specularity in our mirror are less than
$0.5\,\mbox{mrad}$.  This performance is at least a factor of
three better than the best atomic mirrors based on evanescent
waves \cite{Asp} and magnetic media \cite{Saba}, making our pulsed
magnetic mirror the smoothest cold atom mirror yet demonstrated.

In addition to the extremely good optical quality, the mirror also
has the property of being adaptive (changing $B_c$ changes the
focal length) and inelastic (changing the pulse duration changes
the vertical impulse imparted to the BEC).  A second experiment
was performed to explicitly take advantage of these aspects of the
mirror. We realized a BEC ``beam expander'', by first expanding
the BEC by reflection from a short radius $(R_{xy}=h)$ mirror, and
then on the second reflection launching \cite{KettICAP} and
collimating it by a longer interaction with a weakly focusing
mirror. The results are shown in Fig.~\ref{BounceSeq} (c). This
process produces a low density, ultracold, condensate fountain,
which might find application, for example, in atomic clocks. It
can also be regarded as a demonstration of delta-kick cooling
\cite{Amm}.

Fig.~\ref{BecRadii} shows that the evolution of a condensate
reflected by the magnetic mirror is well described by classical
physics. This is in contrast to the reflection of condensates by
light sheets \cite{Bongs}, where the reflected condensate develops
a complicated self-interference structure near the turning point
of its trajectory. Fringes occur when the de Broglie wavelength of
the condensate is large at positions where the classical atomic
trajectories cross, a condition which is never satisfied for our
mirror. Of course in many atom-optical applications the absence of
interference fringes is an advantage.

Our two theoretical models do suggest that non-classical evolution
should be visible in the $xy$ plane under certain conditions. In
particular, for the case $R_{xy}=h$ (Fig.~\ref{BounceSeq} (a)) the
density and the ``waist'' size at the tight focus near the
$R_{xy}/2$ point are respectively high enough and small enough
that the effects of atomic interactions and diffraction might be
visible.  The condensate's behavior near this tight waist provides
a sensitive probe of the mirror quality, so it would therefore be
interesting to image the condensate along the trap axis and
compare the results with a full calculation using the
time-dependent 3D Gross-Pitaevskii equation. Another topic for
future investigation is the coherence of the mirror. Since the
trap field is sufficiently well-controlled to produce a condensate
in the first place, it seems very likely that the mirror will
preserve coherence. Nevertheless it would be desirable to
demonstrate this directly, e.g. by splitting a condensate
coherently and then interfering the two reflections.

In conclusion, we have demonstrated adjustable focusing and
reflection of a BEC from a pulsed magnetic mirror, and shown that
the evolution of the reflected condensate is well-described by a
simple theory.  The mirror establishes a new limit for the optical
quality of atom-optical elements. Finally, we have also used the
adaptive and inelastic properties of the mirror to realize an
atomic ``beam expander'' and used it to make a condensate
fountain.

We gratefully acknowledge valuable discussions with Ed Hinds. This
work was supported by the UK EPSRC and the University of Sussex.
ASA was funded by the Commonwealth Scholarship Commission.

\end{document}